\documentclass[10pt]{article}  
\newcommand{\comma}{\, , \; \; }
\newcommand{\period}{\, .}
\newcommand{\eq}{\; = \;}
\newcommand{\sep}{\, , \;\;}

\newcommand{\be}{\begin{equation}}
\newcommand{\bd}{\begin{displaymath}}
\newcommand{\ee}{\end{equation}}
\newcommand{\ed}{\end{displaymath}}
\newcommand{\ba}{\begin{eqnarray}}
\newcommand{\ea}{\end{eqnarray}}

\title{Proof of the determinantal form of the spontaneous magnetization 
of the superintegrable  chiral Potts model}

\author{ R.J. Baxter\\
{\protect \small  Mathematical
Sciences Institute,  The Australian National}\\
{\protect  \small  University,
 Canberra, A.C.T. 0200, Australia}}

\date{\protect \small  2 January 2010}

\begin{document}

%\magnification = \magstep1
%\magnification = 1000

\maketitle
 %%defghijklmnopqrstabcdefghijklmnopqrstabcdefghijklmnopqrstabcdefghijkl

\abstract{The superintegrable chiral Potts model has many resemblances 
to the Ising model, so it is natural to look for algebraic properties
similar to those found for the Ising model by Onsager, Kaufman and 
Yang. The spontaneous magnetization ${\cal M}_r$ can 
be written in terms of a sum over the elements of a 
matrix $S_{r}$. The author conjectured  the form of the elements,
and this conjecture has been verified by Iorgov {\it et al}.
The author also conjectured in 2008 that this sum could
be expressed as a determinant, and has recently evaluated the 
determinant to obtain the known result for ${\cal M}_r$.
Here we prove that the sum and the determinant are indeed
identical expressions.}

 %%defghijklmnopqrstabcdefghijklmnopqrstabcdefghijklmnopqrstabcdefghijkl

\section{Introduction}
There has been considerable progress recently towards the goal of 
finding an algebraic method of calculating the spontaneous 
magnetization
${\cal M}_r$ (the order parameter) of the superintegrable 
$N$-state chiral Potts model. By considering the square lattice of 
finite width
$L$, the author\cite{paper1}-\cite{paper3} showed that it could be 
expressed in terms of a sum over the elements of a 
$2^m$ by $2^{m'}$ matrix $S_{r}$.
In \cite{paper3} he conjectured a formula for these elements as simple
products. He also conjectured in \cite{paper2, paper3} that the sum 
could be written
as the determinant $D_{PQ}$ of an $m$-dimensional 
(or $m'$-dimensional) matrix. These $P, Q, r$ are related by
$Q = P+r$ (mod $N$), and  (for $0 \leq P, Q < N$)
\be \label{defmm}
m = \left[ \frac{(N-1)L-P}{N} \right] \sep 
 m' = \left[ \frac{(N-1)L-Q}{N} \right] \ee
where $[x]$ is the integer part of $x$. Note that these defintions 
imply
\be  \label{resmn}
| m - m' | \leq 1 \period \ee

In a recent paper\cite{paper4}, the author has shown that 
this determinant can be 
written the product (or ratio) of Cauchy-like determinants, so is
also a simple product. Taking the limit $L \rightarrow \infty$, one 
does indeed regain the known 
result\cite{Albertini1989, Baxter2005a,Baxter2005b} 
\be \label{Albert}
{\cal M}_r \eq (1-{k'}^2)^{r(N-r)/2N^2} \period \ee

This still left open the two conjectures, the first being
the product expression for the elements of 
$S_{r}$. The matrix $S_r$ satisfies
two commutation relations. In unpublished work, the author 
had been able to prove 
that the conjectured form did in fact satisfy these relations, 
and from numerical studies for small values of
$N, L$ it appeared that there was only one such solution to these 
linear equations, but  this fell short of a proof. The problem of 
calculating such matrix elements has been studied  more directly
by Au-Yang and 
Perk.\cite{AuYangPerk2010a,AuYangPerk2010b}  Iorgov {\it et al} 
have now given  a proof, and have gone on to 
calculate ${\cal M}_r$ directly.\cite{Iorgov2009}

The second conjecture was that the sum over the elements of 
$S_r$ was the determinant $D_{PQ}$. This is the problem we address 
here. It is a very self-contained problem, being just an algebraic
identity between rational functions of many variables.
It completes the algebraic proof of the formula (\ref{Albert}) 
via the determinant $D_{PQ}$.

We shall refer to papers \cite{paper1} - \cite{paper4} as papers
I, II, III, IV, respectively, and quote their equations
accordingly.

\section{Definitions}

Let $c_1, \ldots, c_m, y_1, \ldots, y_m$ and 
$c'_1, \ldots, c'_n, y'_1, \ldots, y'_n$ be sets of variables,
where
\bd n = m' \ed
and $m, m'$ are the integers mentioned above. In this paper
we do {\em not} use the above definitions (\ref{defmm}), nor the 
integers $N, L, P, Q$. We can take $m, m'$ to both be arbitrary
positive integers, and the $c_i, y_i, c'_i, y'_i$ to be arbitrary
variables.
In this paper we can and do allow $m$ and $m'$ to be arbitrary.
We {\em ignore} the restriction (\ref{resmn}).

 Let $s = \{ s_1, \ldots , s_m \}$ be a set of $m$ integers
with values
\be s_i \eq  0 \; \; {\rm or} \; \; 1 \; \; \; \;  
{\rm for} \; \; 1 \leq i \leq m \period \ee
Similarly, let $s' = \{ s'_1, \ldots , s'_n \}$, where each 
$s'_i = 0$ or $1$.
Set
\be \label{defkappa}
\kappa_s \eq s_1 + \cdots + s_m \sep \kappa_{s'} \eq s'_1 + 
\cdots + s'_n \period \ee

For a given set $s$, let $V$ be 
the set of integers $i$ such that
$s_i= 0$ and $W$ the set such that $s_i = 1$.
Hence, from (\ref{defkappa}), $V$ has $m-\kappa_s$ 
elements, while  $W$  has $\kappa_s$.
Define $V'$, $W'$ similarly for the set $s'$, so
$V'$ has $n-\kappa_{s'}$ elements, while 
$W'$  has $\kappa_{s'}$.
 
Define
\ba \label{defADT}
A_{s,s'} = \prod_{  i \, \in\, W } \prod_{ j \, 
   \in V' } (c_i-c'_j)  &  , &
B_{s,s'} = \prod_{  i \, \in\, V } \prod_{ j \, 
   \in W' } (c_i-c'_j)\comma \nonumber \\
 C_{s}   \eq \prod_{  i \, \in\, W } \prod_{ j \, 
   \in V } (c_j-c_i)   & , &
D_{s'} \eq  \prod_{ \, i \, 
   \in V' } 
\prod_{  j \, \in\, W' \, }(c'_j-c'_i) \period
\nonumber\ea

Then the afore-mentioned matrix $S_r$ has elements
$(S_r)_{s,s'}$ which are proportional to
$A_{s,s'} B_{s,s'} /( C_{s} D_{s'})$ and the sum over
its elements is given in III.3.48 as 
\be \label{defR}
{\cal R}   \eq \sum_{s} \sum_{s'} 
\, y_1^{s_1} y_2 ^{s_2} \cdots y_m^{s_m} 
\left( \frac{  A_{s,s'} B_{s,s'} }
{   C_s D_{s'}  } \right) 
 {y'_1}^{s'_1} {y'_2} ^{s'_2} \cdots 
{y'_n}^{s'_n} \comma \ee
the sum being restricted to $s,s'$ such that 
\be \label{restrict}
\kappa_s = \kappa_{s'} \period \ee

 Now define $a_i, \ldots, a_m , b_1, \ldots b_m$ and 
$a'_i, \ldots, a'_n , b'_1, \ldots b'_n$  by
\be \label{defa}
a_i = \prod_{j=1}^{n} (c_i - c'_j) \sep 
a'_i = \prod_{j=1}^{m} (c'_i - c_j) \comma \ee
\be \label{defb}
 b_i = \prod_{j=1, j \neq i }^m (c_i - c_j) \sep
b'_i = \prod_{j=1, j \neq i }^{n} (c'_i - c'_j) \comma \ee
and let ${\cal B}$ be an $m$ by $n$ matrix, and ${\cal B}'$  an 
$n$ by $m$  matrix,   with elements
\be {\cal B}_{ij} = \frac{a_i}{b_i (c_i-c'_j)} \sep 
{\cal B}'_{ij} = \frac{a'_i}{b'_i (c'_i-c_j)} \period \ee

Also, define an $m$ by $m$ diagonal matrix $Y$ and 
an $n$ by $n$ diagonal matrix $Y'$ by
\be Y_{i,j} \eq y_i \delta_{ij} \sep 
Y'_{i,j} \eq y'_i \delta_{ij} \period \ee
Then the determinant  mentioned above is

\be \label{defD}
{\cal D} \eq D_{PQ} \eq \det [I_m + Y {\cal B} Y' {\cal B}' ]
\comma \ee
where $I_m$ is the identity matrix of dimension $m$.

(The definition (\ref{defD}) is the same as the (II.7.2), (III.4.9),
(IV.1.18). If $f_, f'_i$ are defined by IV.2.31 and $B_{PQ}$ by 
IV.2.29, and
$F, F'$ are the diagonal matrices with elements 
$F_{i,i} = f_i $, $F'_{i,i} = f'_i $, then 
${\cal B} = \epsilon F B_{PQ}{F'}^{-1}$, 
${\cal B}' = -\epsilon F' B_{QP} F^{-1} $ and $\epsilon^2=1$.
We can then observe that  (\ref{defD}) is the same as
IV.1.18.)

We can also write (\ref{defD}) as 
\be \label{defDb}
{\cal D} \eq \det [I_n + Y' {\cal B}'  Y {\cal B} ]\comma \ee
so both $\cal R$ and $\cal D$ are unaltered by 
simultaneously interchanging $m$ with $n$, 
the $c_i$ with the $c'_i$ and the $y_i$ with the $y'_i$. It
follows that without loss of  generality, we can here choose
\be \label{mpm}
n \geq m \period \ee

The expressions ${\cal R}, {\cal D}$ are functions of
$m,n, c_1, \ldots c_m, y_1, \ldots, y_m,
 c'_1, \ldots c'_n, y'_1, \ldots, y'_n$. We shall write
them as ${\cal R}_{mn}, {\cal D}_{mn}$.

%%%%%%%%%%%%%%%%%%%%%%%%%%%%%%%%%%%%%%%%%%%%%%%%%%%%%%%%%%%%%%%%%%%%%%

\section{Proof that ${\cal R}_{mn} = {\cal D}_{mn}$}

Both ${\cal R}_{mn}$ and ${\cal D}_{mn}$ are rational functions 
of $ c_1, \ldots c_m$, $c'_1, \ldots c'_n$. They are symmetric, 
being  unchanged by
simultaneously permuting the $c_i$ and the $y_i$, as well as by
simultaneously permuting the $c'_i$ and the $y'_i$.
We find that they are identical, for all $c_i, y_i, c'_i, y'_i$.

The proof proceeds by recurrence, in the following four steps.

\subsection*{1.  The case $m= 1$}
\subsubsection*{ Calculation of ${\cal R}_{1n}$}  

If  $m=1$ then $s  = \{s_1\}$ and either $s_1 =0$ or $s_1 = 1$. 
In the first case, since $\kappa_s = \kappa_{s'}$, all the $s'_i$ 
must be zero and the sets $W, W'$ are  both empty, so we get a 
contribution to (\ref{defR}) of unity.

In the second case, $s' = \{0,\ldots, 0,1,0, \ldots ,0\}$, with the 
1 in position $r$, for $r=1, \ldots, n$. Then $V$ is empty, 
so $B_{s,s'} = C_s = 1 $, while
\be A_{s,s'} = \prod_{j=1,j \neq r}^{n} (c_1 - c'_j) \sep 
 D_{s'} = \prod_{j=1,j \neq r}^{n} (c'_r - c'_j) \period \ee
{From} (\ref{defR}) it follows  that
\be {\cal R}_{1n} \eq 1 + \sum_{r=1}^{n} y_1 y'_r  
\prod_{j=1, j \neq r}^{n}
\frac{c_1-c'_j}{c'_r-c'_j} \period \ee

\subsubsection*{ Calculation of ${\cal D}_{1n}$}

The RHS of (\ref{defD}) is a determinant of dimension 1, so
\ba {\cal D}_{1n} & = &  1+ Y_{1,1}\sum_{r=1}^{n} 
{{\cal B}}_{1,r} Y'_{r,r} {{\cal B}}'_{r,1} \nonumber \\
& = & 1 -  \frac{a_1 y_1}{b_1} \sum_{r=1}^{n} \frac{a'_r y'_r }
{b'_r  (c_1-c'_r)^2 } \comma  \ea
where $a_i, a'_i,b_i,b'_i$ are defined by (\ref{defa}), (\ref{defb}).
Note that here $b_1 = 1$ .

 We therefore obtain
\be {\cal D}_{1n} \eq 1 + \sum_{r=1}^{n} y_1 y'_r  \prod_{j=1, j 
\neq r}^{n} \frac{c_1-c'_j}{c'_r-c'_j}  \ee
and we see explicitly that
\be {\cal R}_{1n} \eq {\cal D}_{1n} \period \ee

\subsection*{2: Degree of the numerator polynomials}

Consider ${\cal R}_{mn} $ and ${\cal D}_{mn}$ 
as functions of $c_m$. They are both rational functions. We show here 
 that they are both of the form
\be  \label{form}
\frac{{\rm polynomial \; of \; degree } \; ( n -1)}{b_m} \ee

 \subsubsection*{Degree for  ${\cal R}_{mn}$}
First consider the sum ${\cal R}_{mn}$ in (\ref{defR}) as a function 
of $c_m$. Each term is plainly a polynomial divided by $b_m$. 
If $m \in W$, then $s_m = 1$ and the numerator is proportional 
to $A_{s,s'}$. The degree of the numerator is the number of elements 
of
$V'$. The condition (\ref{restrict}) means that $W'$ must have at 
least one element, so $V'$ must have at most $n-1$.  The degree of 
the numerator is therefore at most $n-1$.

If $m \in V$, then the numerator is proportional to
$B_{s,s'}$ and the degree of the numerator is the number of elements 
of $W'$, which from  (\ref{restrict})  is the same as the number of 
elements of $W$. Since $V$ has at least one element, $W$ can have at 
most $m-1$. From (\ref{mpm}), this is at most $n-1$

The sum of all the terms in (\ref{defR})  is therefore a polynomial
in $c_m$ of degree at most $n-1$, divided by $b_m$, as in 
(\ref{form}).

\subsubsection*{Degree for  ${\cal D}_{mn}$}
Now consider the determinant ${\cal D}_{mn}$ in  (\ref{defD}) 
as a function of $c_m$.
At first sight there appear to be poles at $c_m = c'_j$, coming from
${\cal B}_{mj}$. However, they are cancelled by the factor $a_m$.
Similarly, the ones in the element ${\cal B}'_{jm}$ of
the matrix ${\cal B}'$ are cancelled 
by the  factor $a'_j$. So there are no poles at 
$c_m = c'_j$, for any $j$.

There are poles at $c_m = c_i$ (for $1 \leq  i < m$) coming from the 
$b_i, b_m$ factors in ${\cal B}_{ij}, {\cal B}_{mj}$, respectively, 
so there are simple poles in each of the rows $i$ and $m$.
This threatens to create a double pole in the 
determinant ${\cal D}_{mn} $.
However, if  $c_m = c_i$, the rows $i$ and $m$ of the matrix
$(c_m-c_i) {\cal B}$ are equal and opposite. By replacing row $i$ by 
the sum of the two rows (corresponding
to pre-multiplying ${{\cal B}}$ by an elementary matrix), we can 
eliminate the poles in row $i$. Hence there is only a single pole
at  $c_m = c_i$. The determinant
is therefore a polynomial in $c_m$, divided by $b_m$.

To determine the degree of this polynomial, consider the behaviour 
of ${\cal D}_{mn} $  when $c_m \rightarrow \infty$. Then,
writing $c_m$ simply as $c$,
\bd {\cal B}_{ij} \sim {c}^{ -1} \; \; {\rm if } \; \;  i < m  \sep 
 {\cal B}_{ij} \sim {c}^{ n-m} \; \; {\rm if } \; \;  i =  m   \comma  
\ed
\bd {\cal B}'_{ij} \sim c \; \; {\rm if } \; \;  j < m  \sep 
{\cal B}'_{i,j} \sim {1}  \; \; {\rm if } \; \;  j =  m  \period \ed
and hence   the orders of the
elements of the matrix product in (\ref{defD}) are 
given by 
\be Y \, {\cal B} \,  Y' \, {\cal B}'  \sim 
\left( \begin{array}{llcll}
1 & 1 & ...\; \; \;\;\; & 1 & c^{-1} \\
1 & 1 & ... \; \; \;\;\; & 1 & c^{-1} \\
..  & ..  & ... \; \; \;\;\; &  .. & .. \\
1 & 1 & ... \; \; \;\;\; & 1 & c^{-1}  \\
c^{n-m+1} & c^{n-m+1}  & ... \; \; \;\;\; & c^{n-m+1} 
& c^{n-m}  \end{array}
\right) \period \ee
Since $n \geq m$, it follows that ${\cal D}_{mn}$
grows at most as
\be {\cal D}_{mn} \sim   c^{n-m} \ee
as $c \rightarrow \infty$. The numerator polynomial 
in (\ref{form}) is therefore of degree
at most  $n-1$.
This completes the proof of (\ref{form} ).

\subsection*{3. The case $c_m = c'_{n}$}
 \subsubsection*{The sum ${\cal R}_{mn}$}
Consider the case when $c_m \eq c'_{n}$. If $m \in W $ and 
$n \in V'$, then  from (\ref{defADT}) $A_{s,s'} = 0$.
Similarly, If $m \in V $ and 
$n \in W'$, then from  $B_{s,s'} = 0$. So the summand in (\ref{defR})
is zero unless either $m \in V,  n \in V'$, or 
$m \in W,  n \in W'$.

In the first instance, $s_m = s'_{n} = 0$. The $AB/CD$ factor in
(\ref{defR}) is the same as if we take replace $m, n$ by $m-1,n-1$,
respectively, except for a factor
\be  \prod_{i \in W} \frac{c_i - c'_{n}}{c_m - c_i } \; 
\prod_{j \in W'}  \frac{c_m - c'_{j}}{c'_j - c'_{n} } \period \ee
Since $c_m  = c'_{n}$, the factors in the product cancel, except 
for a sign. From (\ref{restrict}),  there as many elements in $W$
as in $W'$, so the sign products also cancel , leaving unity.
Thus this contribution to (\ref{defR}) is exactly that 
obtained by replacing $m, n$ by $m-1,n-1$.

%%defghijklmnopqrstabcdefghijklmnopqrstabcdefghijklmnopqrstabcdefghijkl

In the second instance, $s_m = s'_{n} = 1$. This time  $AB/CD$
has an extra factor
\be  \prod_{j \in V'} \frac{c_m - c'_{j}}{c'_{n} - c'_{j} } \; 
\prod_{i \in V} 
\frac{c_i - c'_{n}}{c_i - c_{m} } \eq 1 \comma \ee
so this contribution to (\ref{defR}) is again that 
obtained by replacing $m, n$ by $m-1,n-1$, except that now there
is an extra factor $y_m y'_{n}$. Adding the two contributions,
we see that 
\be  \label{eqd1}
{\cal R}_{mn}\eq (1+y_m y'_{n})  {\cal R}_{m-1,n-1} \period \ee

 \subsubsection*{The determinant ${\cal D}_{mn}$}
Now look at the determinant (\ref{defD}) when  $c_m \eq c'_{n}$.
Since $a_m$ and $a'_{n}$ both contain the factor
$c_m  - c'_{n}$, the $m$th row of ${\cal B}$ vanishes
except for the element $m,n$, which is
\be {\cal B}_{m,n} \eq  \prod_{j=1}^{n-1} (c_m-c'_j)\left/ 
\prod_{j=1}^{m-1} (c_m- c_j) \right. \period \ee
Similarly, the $n$th row of ${\cal B}'$ vanishes except for
\be {\cal B}'_{n,m} \eq  \prod_{j=1}^{m-1} (c'_n-c_j)\left/ 
\prod_{j=1}^{n-1} (c'_{n}- c'_j) \right. \period \ee
Since $c_m \eq c'_{n}$, we see that 
${\cal B}_{m,n}  {\cal B}'_{n,m} =1$.

It follows that the matrix in (\ref{defD}) has a block-triangular
structure:
\be I_m + Y {\cal B} Y' {\cal B}' \eq 
\left(\begin{array} {cc} {\bf 1 + yby'b'} &  {\bf \cdots} \\
 {\bf 0} &  1 + y_m y'_{n}
\end{array} \right)  \comma \ee
where ${\bf 1 , y, b, y',  b'}$ are the matrices $I_m, Y, 
{\cal B},  Y',  {\cal B}'$ with their last rows and columns 
omitted. Hence
\be  \label{eqd2}
{\cal D}_{m,n} \eq (1+y_m y'_{n})  {\cal D}_{m-1,n-1} \period \ee

\subsubsection*{4. Proof by recurrence}
 The proof now proceeds by recurrence. Suppose ${\cal D}(m-1,n-1) = 
{\cal R}(m-1,n-1) $ for all $c_i,c'_i$. Then from (\ref{eqd1}) 
and  (\ref{eqd2}) it  is  true that ${\cal D}_{m,n} = 
{\cal R}_{m,n} $ when $c_m = c'_{n}$. 
 By symmetry it is also true for $c_m = c'_{j}$ for
$j = 1, \ldots , n$. Thus ${\cal D}_{m,n}- {\cal R}_{m,n}$
is zero for all these $n$ values. But from point 2 above, this 
difference (times $b_m$) is a polynomial in $c_m$ of degree 
$n - 1$. The polynomial must therefore vanish identically, 
so  ${\cal R}_{m,n} = {\cal D}_{m,n}$ for {\em arbitrary} values 
of $c_m$.

Since it is true for $m=1$, it follows that
 \be {\cal R}_{m,n} = {\cal D}_{m,n} \ee
for {\em all} $m, n$.

This proves the second conjecture of \cite{paper3}. 

\section{Summary}
We have proved that the sum ${\cal R}$ over the elements
of the matrix $S_r$ is identical to the determinant
${\cal D}$. In paper  IV we have calculated  ${\cal D}$ 
and hence obtained the spontaneous magnetization ${\cal M}_r$.

The recent work by Irgov {\it et al}\cite{Iorgov2009} proves that 
the elements of the matrix $S_r$ are indeed given by III.3.45, being
proportional to  $A_{s,s'} B_{s,s'} /( C_{s} D_{s'})$, so the 
algebraic calculation of ${\cal M}_r$ is  now complete. 

Further, Irgov {\it et al} go on to
calculate $\cal R$, and hence ${\cal M}_r$,
 directly, thereby avoiding the determinantal 
formulation altogether.  While this last step is efficient, 
it by-passes the author's original 
motivation for this work, which was to 
obtain a derivation  of ${\cal M}_r$ that more closely
resembled the algebraic and combinatorial determinantal calculations
for the Ising model  of Yang, Kac and Ward, Hurst and Green, 
and Montroll, Potts and Ward, \cite{Yang1952} -- \cite{Montroll1963}
 all of whom write the  partition 
function  in terms of a determinant or pfaffian 
(the square root of an anti-symmetric determinant). Indeed, 
the ${\cal D} = D_{PQ}$ of this paper is the immediate
generalization of the Ising  model determinant, as formulated
in I.7.7 in the first paper of this series.

In fact, it would still be illuminating to obtain a simple and direct
derivation of $D_{PQ}$ parallelling Kaufman's spinor operators 
(Clifford algebra) method for the Ising model.\cite{Kaufman1949}

%%%%%%%%%%%%%%%%%%%%%%%%%%%%%%%%%%%%%%%%%%%%%%%%%%%%%%%%%%%%%%%%%%%%%%

 \end{document}